\documentclass[iop,twocolumn]{emulateapj}

\usepackage{amsmath,amssymb,amstext}
\usepackage{xspace}

\usepackage[breaklinks,colorlinks,citecolor=blue,linkcolor=magenta]{hyperref} 

\usepackage[all]{hypcap} 

\usepackage{natbib}
\usepackage{booktabs}
\usepackage{array}
\newcolumntype{L}{>{\centering\arraybackslash}m{1.5cm}}  
\newcolumntype{M}{>{\centering\arraybackslash}m{2.5cm}} 
\newcommand{\rstar}{\ensuremath{R_*}\xspace} 
\newcommand{\rearth}{\ensuremath{R_{\oplus}}\xspace} 
\newcommand{\mearth}{M\ensuremath{_{\oplus}}\xspace}

\shorttitle{Plausible Compositions of the Seven TRAPPIST-1 Planets Using Long-term Dynamical Simulations}
\shortauthors{Quarles et al.} 

\begin{document}

\title{Plausible Compositions of the Seven TRAPPIST-1 Planets Using Long-term Dynamical Simulations}
\author{B. Quarles}
\affil{HL Dodge Department of Physics \& Astronomy, University of Oklahoma, Norman, OK 73019,
USA}
\email{billylquarles@gmail.com}
\author{E. V. Quintana, E. Lopez, J. E. Schlieder, T. Barclay}
\affil{NASA Goddard Space Flight Center, 8800 Greenbelt Rd, Greenbelt, MD 20771, USA}

\begin{abstract}
TRAPPIST-1 is a nearby ultra-cool dwarf that is host to a remarkable planetary system consisting of seven transiting planets. The orbital properties and radii of the planets have been well-constrained, and recently the masses of the inner six planets have been measured with additional ground and space-based photometric observations. Large uncertainties in these mass measurements have prevented a robust analysis of the planetary compositions.  Here we perform many thousands of N-body dynamical simulations with planet properties perturbed from the observed values and identify those that are stable for millions of years. This allows us to identify self-consistent orbital solutions that can be used in future studies. From our range of dynamical masses, we find that most of the planets are consistent with an Earth-like composition, where TRAPPIST-1f is likely to have a volatile-rich envelope. 
\end{abstract}

\maketitle

\section{Introduction} \label{intro}
Recently, seven roughly Earth-sized planets have been discovered transiting a nearby ultra-cool dwarf star, TRAPPIST-1 \citep{Gillon2017,Luger2017}.  This system has revealed that planet formation can indeed occur efficiently around the lowest mass stars, which when combined with the large number of planets, make this system of great interest to the astronomical community.  In this discovery, precise planetary properties were determined using photometric data from ground-based observatories, the Spitzer Space Telescope, and the Kepler Space Telescope in its two-reaction wheel mission \citep[K2][]{Howell2014} as part of Campaign 12.  

As a result, the orbital periods, planetary radii, impact parameters, and sky-plane inclinations are well characterized.  \citeauthor{Gillon2017} noted that periods between adjacent planets followed roughly integer multiples in period ratios, indicating that mean motion resonances could play a significant role in the long-term evolution of the system.  Moreover, models based on the analysis of transit timing variations (TTVs) showed that the system could become unstable on relatively short timescale ($\sim$0.5 Myr), where simulations including tidal dissipation did not greatly enhance the long-term stability.  The TTV analysis also provided upper limits on the eccentricity of the planets, but these limits are degenerate with the planetary mass estimates.

\citeauthor{Luger2017}, using specifically the K2 data \citep{Howell2014}, detected the seventh planet and identified that resonant chains likely exist between members of the planetary system.  This suggests a fairly quiescent formation path where excitations in eccentricity are quenched through interactions in the gas phase of the protoplanetary disk.  They find through Fourier analysis of starspots from the K2 data that the host star likely has a rotation period of $\sim$3 days and an age of 3--8 Gyr.  The resonant chain therefore needs to persist on Gyr timescales so the range of possible planetary parameters is limited.

By identifying the planetary parameters that are stable for at least 1 Myr timescales, masses (and therefore densities and composition estimates) for the seven TRAPPIST-1 planets can be estimated.  This becomes important when modeling the evolution of the system up to the stellar age as tidal interactions will become more important.

In this paper we present N-body simulations that we use to ascertain which planet parameters can produce stable solutions for 1 Myr timescales. This allows us to infer the physical properties of the planets.

\section{Methods} \label{methods}
\subsection{Numerical Setup}
\cite{Luger2017} and \cite{Gillon2017} detail many of the initial planetary properties; we use their properties and summarize them in Table \ref{tab:ICs}.  In order to perform our numerical simulations, we use a modified version of the \texttt{mercury6} integration package \citep{Chambers1999} that has been optimized to identify collisions between planets, encounters with the host star, and scattering events as conditions to stop a simulation.  We use the hybrid integration scheme to efficiently evaluate the orbital evolution with an integration step that is 5\% the orbital period of TRAPPIST-1b.  As an additional check on our results, we also use the \textit{REBOUND} integration package \citep{Rein2012} for a subset of runs implementing the \texttt{WHFAST} integrator \citep{Rein2015}.

\begin{figure*}
\centering
\plotone{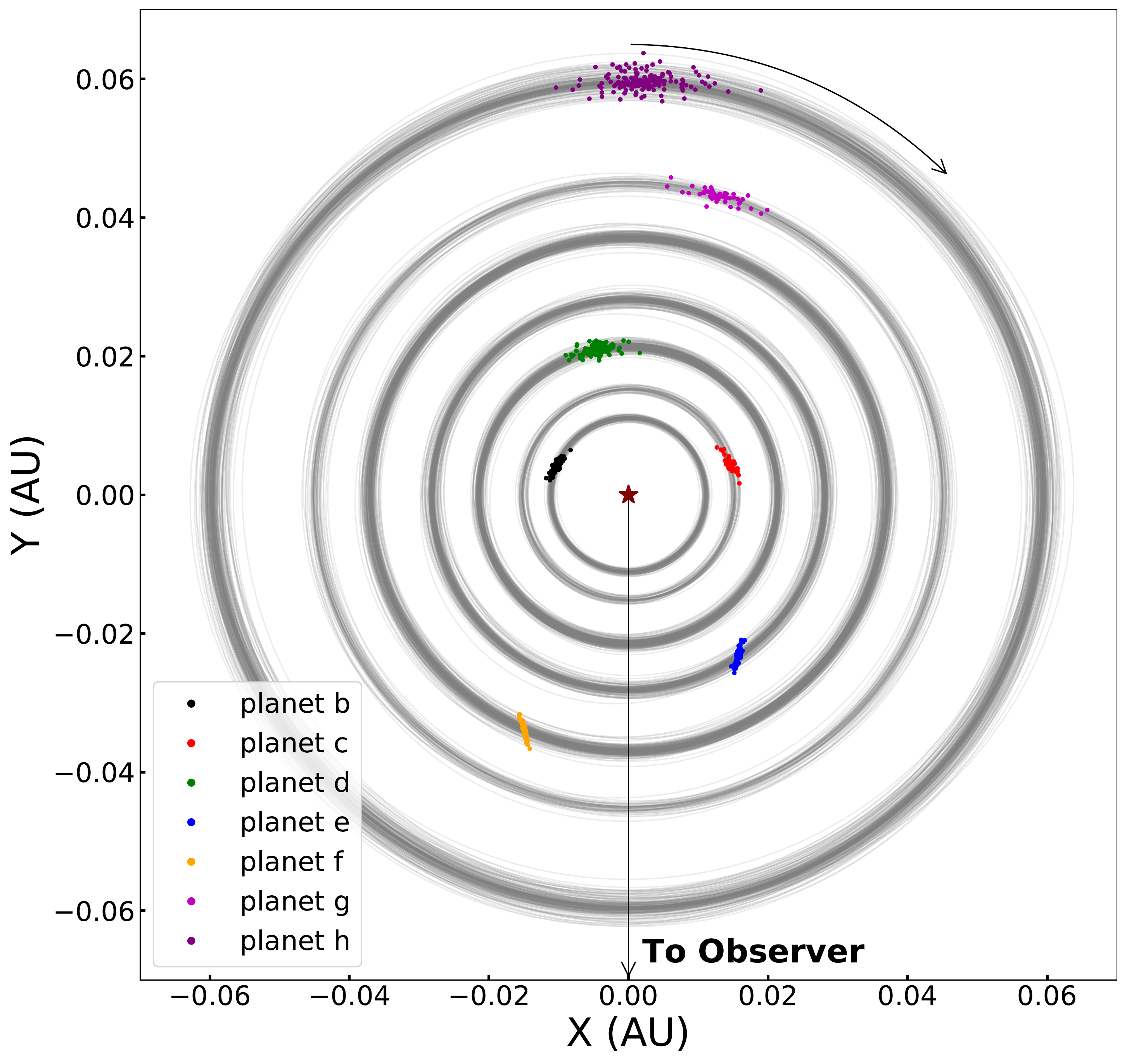}
\caption{The starting locations of our samples that survive for 1 million years relative to the observer.  The spread of colored points results from the observational uncertainties in light curve parameters (i.e., impact parameter, scale parameter, and sky-plane inclination) that are used to deduce the phase at $t_o$ = 2457672 BJD.  The gray curves show the variation in the initial orbits.  The curved arrow denotes the direction that the planets orbit in this top-down view.}
\label{fig:ICs}
\end{figure*}

The initial states for each of the planets are determined using the properties given in Table \ref{tab:ICs}, {where each value is drawn from a Gaussian distribution assuming the \citeauthor{Gillon2017} and \citeauthor{Luger2017} results have 1-sigma uncertainties unless stated otherwise.}  We use draws from the distributions on the planetary impact parameter $b$, the planetary eccentricity $e$, the scale $a/R_\star$, and the sky-plane inclination $i_{sky}$ to deduce the argument of periastron $\omega$ and mean anomaly $M$ at a common epoch, $t_0= 2457672$ BJD \citep{Murray2000,Winn2010}.  As the mutual inclinations between the planets are small, we set the initial longitude of ascending node $\Omega$ of all the planets to zero {(See Figure \ref{fig:ICs})}.

In order to find the state of the system at a common epoch based on our chosen initial parameters, we use the \texttt{WHFAST} integrator to evolve each planet, treated as a test particle, from its transit epoch along its orbit up to the common epoch $t_0$.  Our sample of initial states are then evolved in parallel up to 1 Myr of simulation time.  We continue this process until at least 5,000 samples survive up to our assumed threshold.  Continuing our integrations beyond 1 Myr can be problematic because long-term effects due to General Relativity and/or tides may become important, and we leave this for future studies.

\section{Results and Discussion} \label{results}
\subsection{Distributions of the Survivors}
Our numerical simulations resulted in a wide range of outcomes, where we are interested in them as a group and do not prescribe to have found unique orbital solutions.  In total we performed {18,527 trials and 28.2\%} of those were found to be stable for 1 Myr.  A summary of parameters pertaining to the dynamical stability are presented in Table \ref{tab:final_dist}.

The estimate for the planetary masses in Table \ref{tab:ICs} were determined through a TTV analysis \citep{Gillon2017}, where our masses (in Table \ref{tab:final_dist}) are the result of 1 Myr stability simulations.  As a result, we find that the nominal values from the distributions to be consistent, but the uncertainties to be different.  This difference can be explained, in part, by the additional planet in this work (TRAPPIST-1h) and that some solutions fit the data fairly well, but result in an unstable long-term orbital solution {usually shrinking the range at 1-sigma}.

The numerical scheme within \texttt{mercury} that resolves the collisions is particularly sensitive to the assumed radius of each planet.  Our samples made draws in the range of planetary radii from the discovery papers \citep{Gillon2017,Luger2017}, where those simulations that survive up to 1 Myr  are consistent with the observations.  The planetary eccentricities of stable configurations also tended to smaller values.  Combining our distributions of planetary masses and radii, we provide the resulting distribution in mean planetary density.  These values are generally consistent with those of \cite{Gillon2017}.

The tendency of our results towards lower eccentricities is strongly due to relatively small dynamical spacing of the planets.  We provide a measure of the dynamical spacing through the mutual Hill radius ($R_{H}^{i,i+1}$) {using the semimajor axis ($a_i,a_{i+1}$) and our assumed masses ($m_i,m_{i+1}$) of adjacent planets \citep{Chambers1996} as follows:}

\begin{align}
R_{H}^{i,i+1} &= \frac{(a_i + a_{i+1})}{2}\left(\frac{m_i+m_{i+1}}{3M_\star}\right)^{1/3} 
\end{align}
and
\begin{align}
\beta^{i,i+1} &= \frac{a_{i+1} - a_i}{R_{H}^{i,i+1}}.
\end{align}

Previous studies \citep{Chambers1996,Smith2009,Pu2015,Obertas2017} have found that long-term stable solutions of 3--5 planets require spacings with $\beta^{i,i+1}>10$ mutual Hill radii and similar spacings for more than 5 planets.  This is the case for most of the TRAPPIST-1 planets, \textit{except} for the spacing between planets f and g ($\beta^{f,g} = 6.61^{+0.97}_{-0.66}$).  Our simulations probe a range of masses, so that the upper limits given in $\beta^{i,i+1}$ represent those realizations where the sum of the mass for adjacent pairs is low and vice versa.

In order to explain how planets f and g can remain stable, we look to the initial mean longitudes ($\lambda = \omega + M$) of the stable solutions because tighter spacings are possible if the relative phasing between the planets are appropriate and permit long-term stability, akin to the phasing between Neptune and Pluto within the solar system.  {We find the relative phase between planets f and g to be $\sim$169$^\circ$ -- 172$^\circ$} at our epoch, which is nearly out-of-phase.  {For long-term stable orbits to be possible some resonance phenomena may be responsible, but not required.}  \cite{Luger2017} showed resonant chains, or Laplace resonances, to be active within the system {that depend on the initial mean longitude of a planet with its closes neighbors and \cite{Tamayo2017} have identified stable configurations within the resonance for long timescales.}  We identify the resonant argument $\phi^{i-1,i,i+1}$ of our surviving population in Table \ref{tab:final_dist}.  \cite{Luger2017} reported values for the resonant argument where transits occur at a specific phase angle, planets are assumed to have low eccentricities ($\sim$0.01), and within -180$^\circ$ -- 180$^\circ$.  Our values do not make these assumptions and are within 0$^\circ$ -- 360$^\circ$.

\subsection{Comparison to Other Compact Systems}

The most iconic compact configuration within the solar system is the Galilean moons in orbit around Jupiter, which have dynamical spacings of $\sim$14--16 mutual Hill radii.  Other compact systems have been discovered within the Kepler mission, such as Kepler-11 \citep{Lissauer2011,Lissauer2013}, Kepler-186 \citep{Quintana2014}, and Kepler-223 \citep{Mills2016}.  The dynamical separations of the Kepler-11 planets are typically between 8.5 -- 16.5 mutual Hill radii between each pair of planets, where the Kepler-186 planets are all $>$14 mutual Hill radii apart.  We note that the definition of the mutual Hill radius takes the stellar mass into account in order to make appropriate comparisons.

Recently, \cite{Mills2016} uncovered a resonant chain of sub-Neptune transiting planets, where the dynamical separations are $>$8.5 mutual Hill radii between the planets.  This makes the TRAPPIST-1 system especially unique as we find the dynamical spacings to be $\sim$6.56--12.11, which is much tighter than these other systems and may be common among other compact systems orbiting M-dwarfs discovered in the upcoming TESS mission \citep{Ricker2014}. While there are a limited number of similar compact systems available to make a robust comparison, discoveries of such configurations are important to better understand the processes of planet formation and evolution.

\subsection{Possible Compositions of the TRAPPIST-1 Planets}
\begin{figure*}
\centering
\plotone{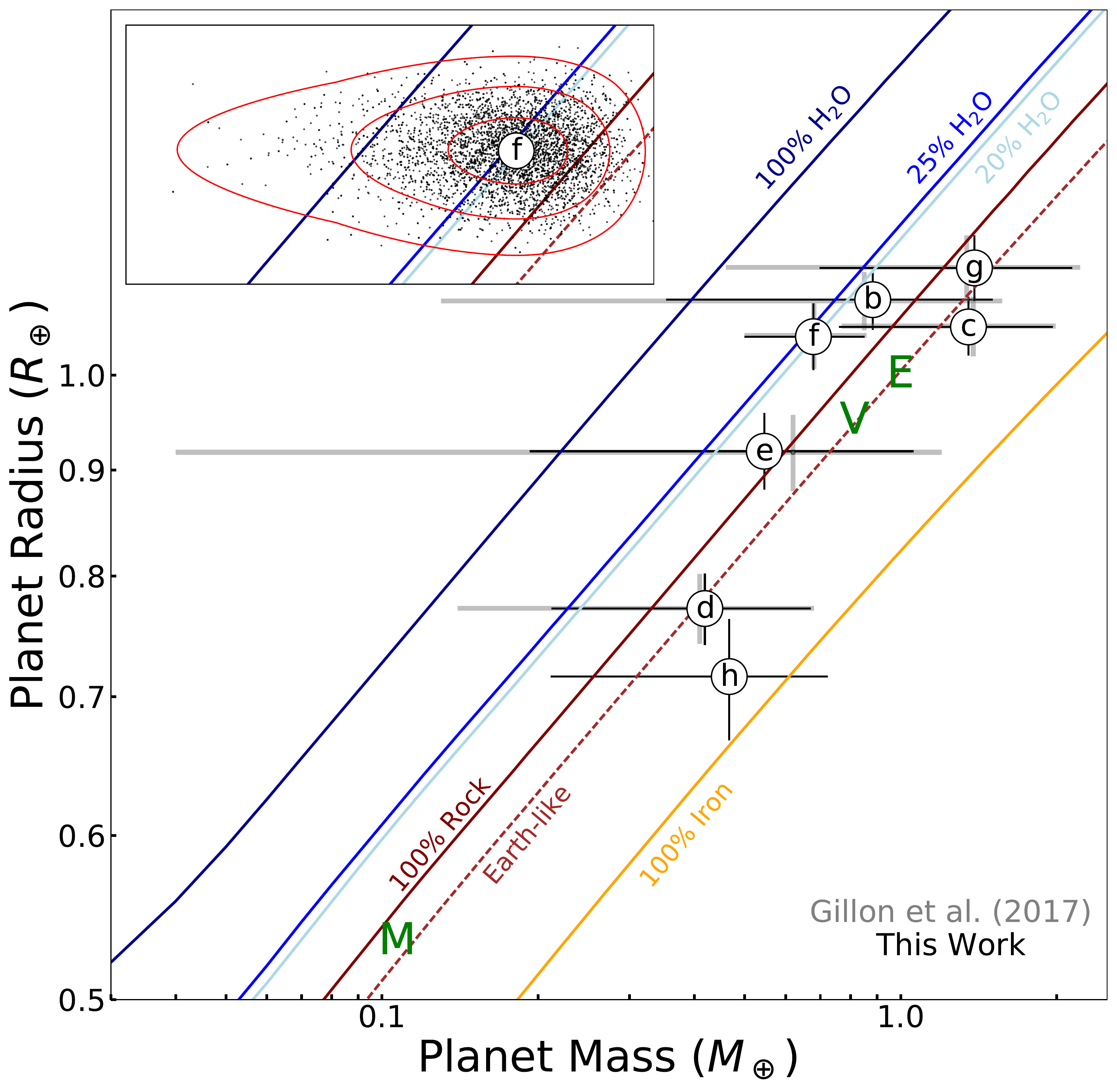}
\caption{Planetary Mass-Radius diagram using the 1 sigma mass and radius uncertainties derived from our simulations.  The TRAPPIST-1 planets are identified by letter, where the errorbars (black \& gray) denote the ranges from this work and \cite{Gillon2017}, respectively.  The inset in the top-left shows the distribution of survivors for planet f, where the contours given in red enclose 66\%, 95\%, and 99.7\% of the distribution.  The terrestrial planets (Venus, Earth, and Mars) have been denoted in green for comparison.
The solid and dashed curves meanwhile show predicted mass-radius relations from \citet{Zeng2016} for planets with different compositions.} 

\label{fig:mr}
\end{figure*}

By combining our new constraints on the masses of the TRAPPIST-1 planets with the previously measured radii, we can set constraints on their possible bulk compositions. Figure \ref{fig:mr} shows the TRAPPIST-1 planets against theoretical mass-radius curves for different bulk planetary compositions from \citet{Zeng2016}. Of the seven planets in the system, six are consistent with having Earth-like or pure rocky bulk compositions, although generally speaking the mass uncertainties are large enough that the planets could also have a significant fraction of their mass in a large volatile envelope formed from water or other volatile ices.

However, one planet in particular stands out in Figure \ref{fig:mr}. Unlike the other six planets, our constraints on the mass of TRAPPIST-1f, which is also the best constrained of the planets in this system, suggest that it is likely inconsistent with a bare rocky composition. Out of nearly {5200} dynamical simulations that survived for 1 Myr, {just over 11\%, found a mass for TRAPPIST-1f that was large enough to cross the iron-free pure rock curve, and only $\sim$1.5\% of the simulations crossed the Earth-like curve.} We find that TRAPPIST-1f is best fit with a massive water envelope comprising $\sim$20\% of the planet's total mass. This is a key conclusion for assessing the possible habitability of TRAPPIST-1f as a planet with a massive water envelope that cannot have liquid water on its surface. 

Using the H2O-REOS equation of state for water \citep{French2009,Nettelmann2010} and thermal evolution models of \citet{Lopez2014}, we find that even at an age of 8 Gyr the temperature at the bottom of such an envelope will be $\gtrsim$1400K and the pressure will reach $\approx$130 kbar. For comparison, the pressure in the deepest parts of Earth's oceans is $\approx$1 kbar. Moreover, these calculations don't include the possibility of significant tidal heating from planet-planet interactions, which could raise the interior temperature even higher. At such a high pressure and temperature, water will be far beyond the triple point and far too hot for high pressure ices like ice VII and X. Instead, it will exist as a high pressure molecular fluid, much like the deep interiors of Neptune and Uranus \citep{Fortney2011,Nettelmann2011}. Therefore, liquid water would likely only exist as clouds near the top of TRAPPIST-1f's atmosphere and our results suggest that it is no more likely to be habitable than any other gas or ice-giant with water clouds in its atmosphere.

The other possible explanation for the low apparent density of TRAPPIST-1f is that it harbors a modest low-metallicity hydrogen dominated envelope. Again using the thermal evolution models of \citet{Lopez2014}, with a solar composition H+He envelope atop an Earth-like rocky core, we find that the best fit mass and radius of TRAPPIST-1f can be matched if it has modest gaseous envelope reaching a pressure of $\approx$200 bar and comprising $\approx$0.03\% of the planet's total mass. We believe this explanation is unlikely, however, since any such hydrogen-dominated atmosphere on TRAPPIST-1f will likely be unstable to X-ray and extreme UV driven photo-evaporative atmospheric escape \citep{Lammer2003}. Recent X-ray observations taken with XMM-Newton by \citet{Wheatley2017} show that TRAPPIST-1 is active in X-rays with $L_X/L_{bol} = 2-4\times10^{-4}$ and an estimated total XUV emission of $L_{XUV}/L_{bol} = 6-9\times10^{-4}$. The X-ray luminosity is similar to the quiet Sun, despite TRAPPIST-1's total luminosity being $\sim$2000$\times$ smaller. The measured X-ray luminosity of TRAPPIST-1 is also comparable to earlier-type, more massive M Dwarfs \citep{Wheatley2017} but much larger than high-mass Brown Dwarfs \citep{Berger2010,Williams2014}. Using the photo-evaporation models of \citet{Lopez2016} we find that a 200 bar H/He atmosphere on TRAPPIST-1f would be lost in only $\sim$10 Myr. Likewise, scaling from recent results of \citet{Bolmont2017}, which specifically examined photo-evaporation from TRAPPIST-1b, yields similar results despite their assumption of a total XUV luminosity nearly two orders of magnitude lower than estimated by \citet{Wheatley2017}.  {\cite{Bourrier2017} also find that total XUV irradiation could be strong enough to strip the atmospheres of the inner planets in a few billions years.}
 
These results strongly motivate the need for additional transits to further refine the planet masses, especially for TRAPPIST-1e and 1g. Currently, TRAPPIST-1e appears to be the best prospect for habitability in the system (assuming a putative atmosphere can somehow survive the high-energy emission of the star), however, the mass uncertainties are quite large and allow for anything from a Mercury-like iron-rich rocky planet to one that is completely dominated by water. Reducing this uncertainty will be a critical first step to any future efforts to characterize these planets with transmission spectroscopy.

\subsection{TRAPPIST-1 in Context: Planetary Systems Around Ultra-Cool Dwarfs}
The discovery and characterization of the planets in the very 
low-mass TRAPPIST-1 system represents a paradigm shift in our 
understanding of planet formation and evolution. The system 
provides a direct probe into the limiting conditions for 
planet formation at the bottom of the main sequence and is a 
benchmark for formation theory and modeling in very-low mass 
disks \citep[e.g.][]{Ormel2017, Ogihara2009, Payne2007, Raymond2007}. Very few comparable systems 
exist, a NASA Exoplanet Archive \citep{Akeson2013} search for 
similarly small planets with periods $<$50 days  
orbiting like host stars ($\mathrm{T_{eff} \lesssim 3000 K~and~M_* \lesssim 0.1 M_{\odot}}$) 
returns only two other systems; 
Proxima Cen \citep{Anglada2016} and Kepler-42 
\citep{Muirhead2012}. 

Surveys for infrared excesses associated with primordial disks 
orbiting young stars across the mass spectrum led to estimates of gas dispersal timescales from $\mathrm{\sim1 - 10~Myr}$ 
\citep[e.g.][]{Haisch2001}. However, discoveries of some young, 
very low-mass stars hosting gaseous accretion 
disks at ages as old as $\sim$20 - 50 Myr \citep[][and references 
therein]{Murphy2017, Silverberg2016, Rodriguez2014} force a 
reconsideration of these timescales and may shed light on 
the circumstances of the TRAPPIST-1 system's formation and 
evolution. Possible consequences of long term disk 
retention are eccentricity dampening and convergent 
migration \citep{Ogihara2009}, potentially yielding compact 
systems in resonant chains like TRAPPIST-1.

Additionally, H and He 
gas may be accreted onto planetary cores for long periods of 
time in these long-lived disks. This possibility is intriguing in 
light of statistical results from gravitational microlensing surveys which 
suggest that $\sim$Neptune mass planets are the most common beyond 
the H$_2$O ice-line \citep{Shvartzvald2016, Suzuki2016} and support that 
the majority of microlens planet host stars are M dwarfs 
\citep{Zhu2017}. Microlensing detections also include several systems with
$\sim$Earth to Neptune mass planets on $\sim$1 AU orbits around 
ultra-cool dwarfs \citep[][and references therein]{Nagakane2017, Shvartzvald2017}. 
These statistics and individual examples provide interesting hypothetical cases for dynamical stability simulations of the TRAPPIST-1 system that 
include additional planets at wider separations. 

The future of planets orbiting ultra-cool dwarfs is promising. 
Ongoing observations of these stars from the ground and space 
will likely provide more analogs to the TRAPPIST-1 system. Near 
future facilities, like precision IR Doppler spectrographs 
\citep[e.g.~iLocater,][]{Crepp2016}, the TESS mission 
\citep{Ricker2014}, and new high-contrast imaging capabilities (JWST, 30m class telescopes) will provide increased sensitivity and wide sky coverage for additional discoveries. The new generation of ground based microlensing surveys \citep{Henderson2014} and the planned microlensing survey of the WFIRST mission \citep{Spergel2013} provide increased observing cadence and sensitivity to facilitate further detections of wide separation ultra-cool dwarf planets 
and may lead to the discovery of a substantial population \citep{Nagakane2017}. Such results, when combined with complementary results from RV, transit, and direct imaging surveys, will eventually enable a comprehensive study of ultra-cool dwarf planetary 
demographics, formation, and evolution.

\section{Conclusions} \label{conclusions}
We have performed numerical simulations of the TRAPPIST-1 system that include all 7 known planets. We perturbed the planet parameters from those reported in the two discovery papers to explore which physical properties yield stable orbits. This allows us to further refine the orbital properties of the planet, and identify plausible compositions for the planets.  We find that 6 of the 7 planets are consistent, within errors, with an Earth-like composition.  The exception is planet f, which is likely to have a volatile-rich envelope, has the tightest constraints due to its TTVs \citep{Gillon2017}, and has a relatively small dynamical separation in mutual Hill radii with planet g.

\cite{Gillon2017} noted that the system was likely to be meta-stable, where a future instability event (i.e., collision) could drastically change the observed planetary architecture.  However, we find that including planet h increases the prospects for stability, likely due to its participation in the resonant chain \citep{Luger2017}.  Stable simulations can be chosen by using the limits on the mean longitude $\lambda_i$ of each planet and the dynamical spacing $\beta^{i,i+1}$ in mutual Hill radii as informative priors until further observations provide better constraints.  We have extended a subset of runs to 10 Myr in this way and found them to stable.

A natural extension to this problem is to consider external perturbers (i.e., longer-period planets) which have been detected around some ultra-cool dwarfs by microlensing and may be common beyond the ice-line.  We have evaluated a subset of our stable configurations including an outer Neptune (15 \mearth) on a nearly coplanar, circular orbit.  We find that the inner compact system remains stable with an outer Neptune beyond 0.37 AU.  This may still be close enough to affect the observed TTVs and should only be considered as a lower limit.

\begin{table*}[h]
\normalsize
\caption{Initial Planet Properties}
\centering
\begin{tabular}{p{2cm} p{2cm} p{2cm} p{2cm} p{2cm} p{2cm} p{2cm} p{2cm}}
 \toprule
  & Planet b & Planet c & Planet d & Planet e & Planet f & Planet g & Planet h\\  
 \midrule
Period (d) & 1.51087081 & 2.4218233 & 4.049610 & 6.099615 & 9.206690 & 12.35294 &18.764 \\ 
Radius (\rearth) & 1.086$^{+0.035}_{-0.035}$ & 1.056$^{+0.035}_{-0.035}$ & 0.772$^{+0.030}_{-0.030}$ & 0.918$^{+0.039}_{-0.039}$ & 1.045$^{+0.038}_{-0.038}$ & 1.127$^{+0.041}_{-0.041}$ & 0.715$^{+0.047}_{-0.043}$ \\
$e$ & $< 0.081$ & $< 0.083$ & $< 0.070$ & $< 0.085$ & $< 0.063$ & $< 0.061$ & $< 0.061$ \\
$b$ (\rstar) &  0.126$^{+0.092}_{-0.078}$ & 0.161$^{+0.076}_{-0.084}$ & 0.170$^{+0.011}_{-0.01}$ & 0.120$^{+0.011}_{-0.011}$ & 0.382$^{+0.035}_{-0.035}$ & 0.421$^{+0.031}_{-0.031}$ & 0.26$^{+0.14}_{-0.16}$ \\
Scale (a/\rstar) & 20.50$^{+0.16}_{-0.31}$ & 28.08$^{+0.22}_{-0.42}$ & 39.55$^{+0.30}_{-0.59}$ & 51.97$^{+0.40}_{-0.77}$ & 68.4$^{+0.5}_{-1.0}$ & 83.2$^{+0.6}_{-1.2}$ & 114$^{+5}_{-5}$ \\
$i_{sky}$ (deg.) & 89.65$^{+0.22}_{-0.27}$ & 89.67$^{+0.17}_{-0.17}$ & 89.75$^{+0.16}_{-0.16}$ & 89.86$^{+0.10}_{-0.12}$ & 89.68$^{+0.0304}_{-0.0304}$ & 89.71$^{+0.025}_{-0.025}$ & 89.8$^{+0.3}_{-0.3}$ \\
Mass (M$_\oplus$) & 0.85$^{+0.72}_{-0.72}$ & 1.38$^{+0.61}_{-0.61}$ & 0.41$^{+0.27}_{-0.27}$ & 0.62$^{+0.58}_{-0.58}$ & 0.68$^{+0.18}_{-0.18}$ & 1.34$^{+0.88}_{-0.88}$ & 0.5$^{+0.4}_{-0.4}$ \\
 \bottomrule
\end{tabular}
\label{tab:ICs}
\end{table*}

\begin{table*}[h]
\normalsize
\caption{Properties of Stable Configurations}
\centering
\begin{tabular}{p{2.5cm} p{2cm} p{2cm} p{2cm} p{2cm} p{2cm} p{2cm} p{2cm}}
 \toprule
  & Planet b & Planet c & Planet d & Planet e & Planet f & Planet g & Planet h\\  
 \midrule
Mass (M$_\oplus$) & 0.88$^{+0.62}_{-0.53}$ & 1.35$^{+0.61}_{-0.59}$ & 0.42$^{+0.25}_{-0.21}$ & 0.55$^{+0.51}_{-0.35}$ & 0.68$^{+0.17}_{-0.18}$ & 1.39$^{+0.76}_{-0.69}$ & 0.47$^{+0.26}_{-0.26}$ \\
Radius (\rearth) & 1.087$^{+0.033}_{-0.036}$ & 1.055$^{+0.036}_{-0.033}$ & 0.772$^{+0.031}_{-0.031}$ & 0.919$^{+0.040}_{-0.038}$ & 1.044$^{+0.039}_{-0.038}$ & 1.126$^{+0.041}_{-0.040}$ & 0.716$^{+0.047}_{-0.049}$ \\
Density ($\rho_\oplus$) & 0.68$^{+0.51}_{-0.41}$ & 1.14$^{+0.55}_{-0.50}$ & 0.92$^{+0.57}_{-0.45}$ & 0.70$^{+0.67}_{-0.46}$ & 0.59$^{+0.18}_{-0.16}$ & 0.96$^{+0.55}_{-0.47}$ & 1.26$^{+0.76}_{-0.71}$ \\
$e$ & $< 0.046$ & $< 0.040$ & $< 0.036$ & $< 0.038$ & $< 0.033$ & $< 0.027$ & $< 0.036$ \\
$\lambda$ (deg.) & 202.38$^{+2.29}_{-2.45}$ & 343.20$^{+2.00}_{-2.16}$ & 257.80$^{+1.25}_{-1.26}$ & 56.23$^{+1.72}_{-1.51}$ & 113.56$^{+1.18}_{-1.90}$ & 285.77$^{+1.23}_{-1.36}$ & 272.39$^{+1.65}_{-1.65}$ \\
\midrule
\midrule
$\phi^{i-1,i,i+1}$ (deg.) & & 182$^{+13}_{-12}$ & 42$^{+6}_{-6}$ & 215$^{+12}_{-11}$ &  287$^{+6}_{-7}$ & 174$^{+4}_{-4}$ & \\
Spacing  ($\beta^{i,i+1}$) & 10.26$^{+1.47}_{-1.05}$ & 12.07$^{+1.85}_{-1.22}$ & 11.68$^{+2.24}_{-1.55}$ & 10.95$^{+1.42}_{-1.20}$ & 6.61$^{+0.95}_{-0.65}$ & 9.72$^{+1.69}_{-1.09}$ & -- \\
 \bottomrule

\end{tabular}

\label{tab:final_dist}
\end{table*}

\begin{table*}[h]
\normalsize
\caption{Orbital State of a Stable Configuration for 10 Myr}
\centering
\begin{tabular}{p{2.5cm} p{2cm} p{2cm} p{2cm} p{2cm} p{2cm} p{2cm} p{2cm}}
 \toprule
  & $a$ & $e$ & $i_{sky}$ & $\omega$ & $M$ & mass & density\\  
  & (AU) &  & (deg.) & (deg.) & (deg.) & (\mearth) & ($\rho_\oplus$) \\
 \midrule
planet b	&	0.01111	&	0.03093	&	89.45583	&	352.19383	&	206.68788	&	0.85909	&	0.58661	 \\
planet c	&	0.01522	&	0.00136	&	89.27847	&	104.19423	&	239.05472	&	1.21375	&	1.12481	 \\
planet d	&	0.02144	&	0.02416	&	89.78165	&	86.04894	&	171.55443	&	0.27054	&	0.57540	 \\
planet e	&	0.02818	&	0.03220	&	89.92784	&	232.90168	&	185.58118	&	0.39366	&	0.61447	 \\
planet f	&	0.03707	&	0.00029	&	89.66858	&	15.56090	&	98.02423	&	0.75689	&	0.71637	 \\
planet g	&	0.04510	&	0.00781	&	89.66603	&	179.99311	&	106.72365	&	1.17221	&	0.86341	 \\
planet h	&	0.05960	&	0.01663	&	89.36007	&	256.05882	&	16.91593	&	0.33241	&	0.88457	 \\

 \bottomrule

\end{tabular}
\label{tab:stable}
\end{table*}

\acknowledgments{
{The authors thank the anonymous referee for comments and suggestions that improved the quality of the manuscript.} The simulations presented here were performed using the Pleiades Supercomputer provided by the NASA High-End Computing (HEC) Program through the NASA Advanced Supercomputing (NAS) Division at Ames Research Center and at the OU Supercomputing Center for Education \& Research (OSCER) at the University of Oklahoma (OU).
}


\end{document}